	\documentclass[a4paper,11pt]{article}

	\usepackage[nolist]{acronym}	% nolist to define acronyms outside main text
	\usepackage[export]{adjustbox}	% Make objects fit frame dimensions
	\usepackage{booktabs}	% Include toprule
	\usepackage{comment}
	\usepackage{csvsimple}  % Load .csv tables
	\usepackage[italic]{hepnames}
	\usepackage{physics}
	\usepackage{pos}	% Proceedings of Science format
	\usepackage{siunitx}
	\usepackage{subcaption}  % Include subfigures
	\usepackage{tikz}
	\usepackage[compat=1.1.0]{tikz-feynman}

	\usepackage{hyperref}
	\usepackage{cleveref}	% Must be loaded after hyperref!

	\DeclareUnicodeCharacter{2212}{-}	% Define missing unicode character

	\newcommand{\diff}[2]{\text{d}^{#1}#2~}
	\newcommand{\ircut}{\ensuremath{L}}
	\newcommand{\local}{\text{L}}
	
	\newcommand{\ps}{\text{C}}
	\newcommand{\qedc}{\alpha}
	\newcommand{\qedchad}{{\Delta\qedc}_\text{had}}
	\newcommand{\qedcvar}{\Delta\qedc}
	\newcommand{\svpf}{\bar{\Pi}}
	
	\newcommand{\uvcut}{\ensuremath{a}}
	\newcommand{\vpf}{\Pi}
	\newcommand{\weak}{\sin^2\theta_W}
	\newcommand{\weakhad}{(\Delta\weak)_\text{had}}
	\newcommand{\weakvar}{\Delta\weak}
	
	\begin{acronym}
		\acro{bmwc}[BMWc]{Budapest-Marseille-Wuppertal collaboration}
		\acro{bsm}[BSM]{beyond the Standard Model}
		\acro{chpt}[ChPT]{chiral perturbation theory}
		\acro{cls}[CLS]{Coordinated Lattice Simulations}
		\acro{dhmz}[DHMZ]{Davier-Hoecker-Malaescu-Zhang}
		\acro{fse}[FSE]{finite-size effects}
		\acro{gs}[GS]{Gounaris-Sakurai}
		\acro{hp}[HP]{Hansen-Patella}
		\acro{hvp}[HVP]{hadronic vacuum polarisation}
		\acro{ib}[IB]{isospin breaking}
		\acro{ir}[IR]{infrared}
		\acro{knt}[KNT]{Keshavarzi-Nomura-Teubner}
		\acro{local}[\local]{\textit{local}}
		\acro{mll}[MLL]{Meyer-Lellouch-L\"{u}scher}
		\acro{oet}[OET]{one-end trick}
		\acro{onepi}[1PI]{one-particle irreducible}
		\acro{ps}[\ps]{\textit{point-split conserved}}
		\acro{qcd}[QCD]{quantum chromodynamics}
		\acro{qed}[QED]{Quantum Electrodynamics}
		\acro{tmr}[TMR]{time-momentum representation}
		\acro{svpf}[sVPF]{subtracted vacuum polarisation function}
		\acro{vpf}[VPF]{vacuum polarisation function}
	\end{acronym}

\title{The~hadronic running of the electromagnetic~coupling and electroweak~mixing~angle}
\ShortTitle{The hadronic running of $\alpha$ and $\weak$}

\author*[a,b,c]{Teseo~San~Jos\'{e}}
\author*[a,b,c]{Hartmut~Wittig}
\author[d]{Marco~C\`{e}}
\author[e]{Antoine~G\'{e}rardin}
\author[a]{Georg~von~Hippel}
\author[a,b,c]{Harvey~B.~Meyer}
\author[a,b,c]{Kohtaroh~Miura}
\author[a]{Konstantin~Ottnad}
\author[f]{Andreas~Risch}
\author[a]{Jonas~Wilhelm}

\affiliation[a]{
	PRISMA\textsuperscript{+} Cluster of Excellence and Institut für Kernphysik,\\
	Johannes Gutenberg-Universität Mainz, 55099 Mainz, Germany
}
\affiliation[b]{
	Helmholtz-Institut Mainz,\\
	Johannes Gutenberg-Universität Mainz, 55099 Mainz, Germany
}
\affiliation[c]{
	GSI Helmholtzzentrum für Schwerionenforschung,\\
	Planckstraße 1, 64291 Darmstadt, Germany
}
\affiliation[d]{
	Albert Einstein Center for Fundamental Physics (AEC)
	and Institut für Theoretische Physik,\\
	Universität Bern,	Sidlerstrasse 5, 3012 Bern, Switzerland
}
\affiliation[e]{
	Aix-Marseille Université,\\
	Université de Toulon, CNRS, CPT, Marseille, France
}
\affiliation[f]{
	John von Neumann-Institut für Computing NIC,\\
	Deutsches Elektronen-Synchrotron DESY,\\
	Platanenallee 6, 15738 Zeuthen, Germany
}
\emailAdd{msanjosp@uni-mainz.de,wittigh@uni-mainz.de}

\abstract{
  We present results for the hadronic running of the
  electromagnetic coupling and the weak mixing angle from simulations
  of lattice QCD with $N_f=2+1$ flavours of $\order{a}$-improved
  Wilson fermions. Using two different discretisations of the vector
  current, we compute the quark-connected and -disconnected
  contributions to the hadronic vacuum polarisation (HVP) functions
  $\svpf^{\Pphoton\Pphoton}$ and $\svpf^{\PZ\Pphoton}$ for spacelike
  squared momenta $Q^2\leq 7$ $\mathrm{GeV}^2$. Our results are
  extrapolated to the physical point using ensembles at four lattice
  spacings, with pion masses in the range from 130 to 420\,MeV. We
  observe a tension of up to 3.5 standard deviations between our
  lattice results for $\Delta\alpha_{\rm had}^{(5)}(-Q^2)$ and
  estimates based on the \textit{R}-ratio for space-like momenta in
  the range $Q^2=3-7\,\rm GeV^2$. To obtain an estimate for
  $\Delta\alpha_\mathrm{had}^{(5)}(M_Z^2)$, we employ the Euclidean
  split technique. The implications for a comparison with global
  electroweak fits are assessed.
  \acresetall\vspace*{0.5cm}
	\begin{flushright}
  DESY-22-194\\
  	MITP-22-099
	\end{flushright}
}

\FullConference{%
 The 39th International Symposium on Lattice Field Theory, LATTICE2022\\
 Rheinische Friedrich-Wilhelms-Universität Bonn\\
 8th-13th August 2022
}

\begin{document}

\maketitle

\section{Introduction}

Precision physics is one of the main avenues towards the discovery of
particles and interactions \ac{bsm}. In this approach, an improved
experimental determination confronts an equally accurate theoretical
prediction to disentangle well-known phenomena from small, novel
effects.
Electroweak global fits \cite{Zyla:2020zbs} are one particular
consistency test where high precision is required. They constrain the
Higgs boson mass from its loop contributions to well known quantities,
such as the electromagnetic coupling at the $\PZ$-pole mass
$\qedc(M_{\PZ}^2)$. The latter is usually predicted starting at the
Thomson limit and studying the energy dependence, reaching the
$\PZ$-pole eventually. As it turns out, the main source of
uncertainty is the leading hadronic contribution at low energies,
which is conventionally computed invoking the optical theorem and
using experimentally measured
$\sigma\left(\HepProcess{\APelectron\Pelectron\to\text{hadrons}}\right)$
cross section-data
\cite{Davier:2019can,Jegerlehner:2019lxt,Keshavarzi:2019abf}. Lattice
determinations can replace the data-driven approach with an \textit{ab
  initio} calculation, avoiding the complicated structure of
resonances in the time-like region.

Another precision quantity is the weak mixing angle $\weak$, whose
value at low energies is sensitive to BSM physics but is strongly
affected by hadronic uncertainties. In this regime it is accessible in
neutrino-nucleus scattering experiments, atomic parity violation and
parity-violating electron scattering, yet its value is known much less
accurately than the fine-structure constant. The standard theoretical
determination of its leading hadronic contribution employs the same
electron-positron inelastic scattering data, but needs to separate the
contribution from each quark flavour and re-weight it with the
appropriate weak charge. Our alternative approach relying on lattice
\acs{qcd} allows for a better determination because it avoids the
systematic uncertainty due to flavour separation.

Most of the work presented in this contribution has been published in
Ref.\,\cite{Ce:2022eix}, which can be consulted for further details
and a complete set of references.

\section{Lattice setup}

\begin{figure}
	\centering
	\includegraphics[width=0.5\textwidth]{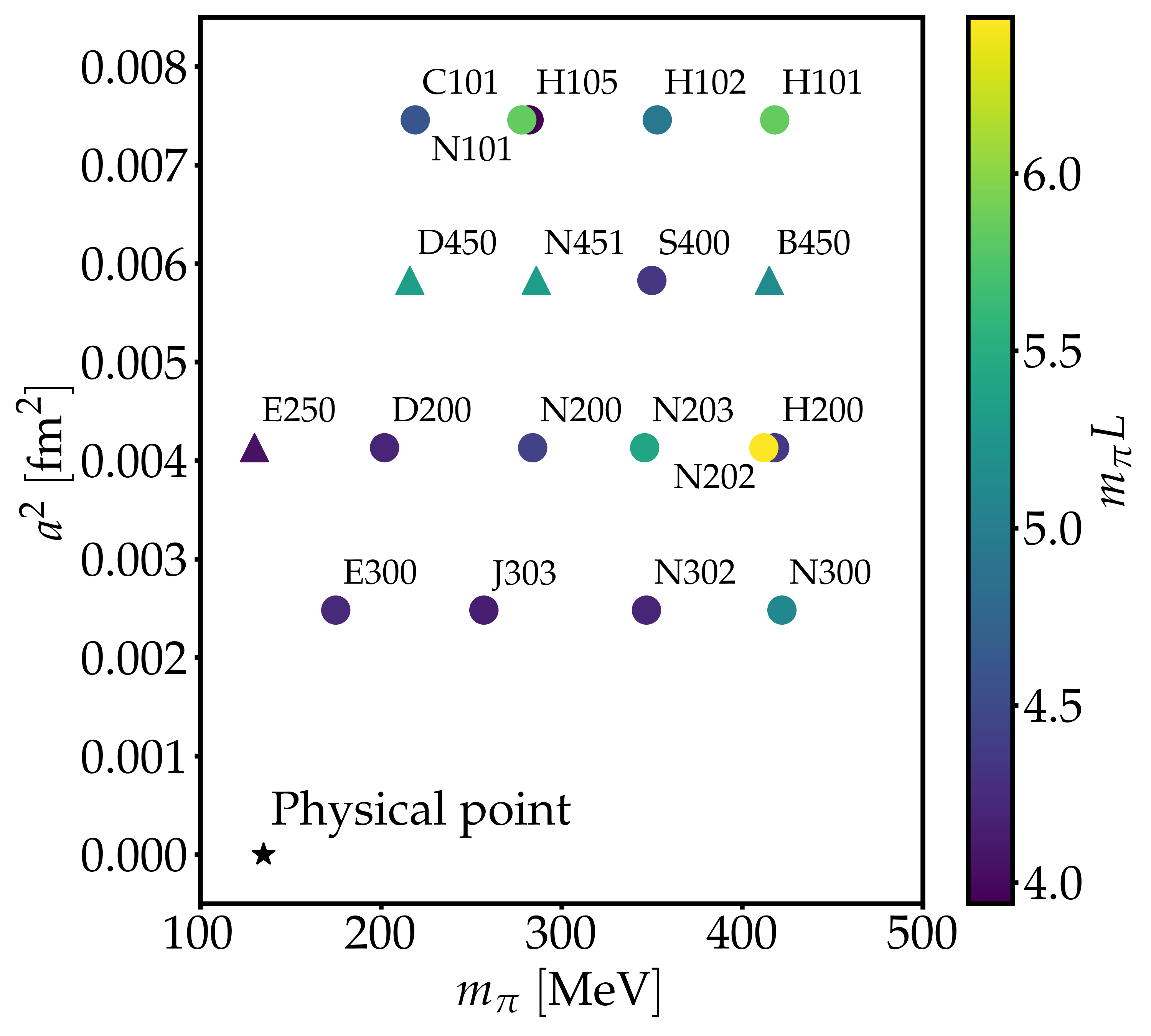}
  \caption{
  	Landscape of \ac{cls} ensembles in terms of the lattice spacing $\uvcut$ and pion mass $m_{\Ppi}$. Triangles (circles) indicate periodic (open) boundary conditions in time.
  }
  \label{fig:ensembles}
\end{figure}

\Cref{fig:ensembles} shows the \ac{cls} ensembles used in our analysis. They include $N_f=2+1$ flavours of non-perturbatively $\order{a}$-improved Wilson fermions, with a tree-level improved L\"{u}scher-Weisz gauge action \cite{Bruno:2014jqa,Bruno:2016plf}.
Besides, we employ the \ac{local} and \ac{ps} discretisations for the quark-bilinears that we analyse in order to better constrain the continuum extrapolation, and
all ensembles lie on the trajectory $m_{\Ppi}^2/2 + m_{\PK}^2 \approx \text{const}$.
The quark loops for the quark-disconnected contribution constitute the most expensive part of the computation, and we calculate them using a variant of the method proposed in \cite{Giusti:2019kff} combining the one-end trick \cite{McNeile:2006bz} with the generalized hopping parameter expansion \cite{Gulpers:2013uca} and hierarchical probing \cite{Stathopoulos:2013aci}. The scale is set using $\sqrt{8t_0} = 0.415(4)(2)~\text{fm}$ \cite{Bruno:2016plf}.

\section{\Acl{tmr}}

The running electromagnetic coupling and weak mixing angle at momentum
transfer $q^2$ are expressed as
\begin{align}
	\label{eq:coupling}
	&\qedc(q^2) = \frac{\qedc(0)}{1-\qedcvar(q^2)},
	&
	&\weak(q^2) = \weak(0)\left(1+\weakvar(q^2)\right),
\end{align}
where $\qedc(0)$ and $\weak(0)$ denote their values in the Thomson
limit. The contributions to the running $\qedcvar(q^2)$ and
$\weakvar(q^2)$ are divided into leptonic and hadronic parts, where
the former can be determined in perturbation theory, and we
concentrate on the latter, which include non-perturbative phenomena at
low momenta. At leading order in the electromagnetic coupling both
quantities, $\qedchad$ and $\weakhad$, can be expressed in terms of
the \acp{vpf} $\svpf^{\Pphoton\Pphoton}$ and $\svpf^{\PZ\Pphoton}$
according to
\begin{align}
	&\qedchad(q^2) = 4\pi\qedc(0) \svpf^{\Pphoton\Pphoton}(q^2),
	&
	\weakhad(q^2) = -\dfrac{4\pi\qedc(0)}{\weak(0)}\svpf^{\PZ\Pphoton}(q^2),
\end{align}
where $\svpf(q^2)=\vpf(q^2)-\vpf(0)$. The subtracted VPF
$\svpf^{\Pphoton\Pphoton}(q^2)$ describes all hadronic \ac{onepi}
diagrams one may introduce in a free photon propagator, while the
corresponding quantity $\svpf^{\PZ\Pphoton}(q^2)$ is relevant for the
mixing between a photon and a \PZ boson. The \acp{vpf}
$\svpf^{\Pphoton\Pphoton}$ and $\svpf^{\PZ\Pphoton}$ can be computed
in lattice QCD for spacelike momentum transfers, $q^2=-Q^2$, using the
\ac{tmr} \cite{2011EPJA...47..148B,2013PhRvD..88e4502F}
\begin{equation}
\begin{gathered}
	\svpf^{\alpha\Pphoton}(-Q^2)
	= \int\diff{}{t}G^{\alpha\Pphoton}(t)K(t,Q^2),
	\\
	G^{\alpha\Pphoton}(t)
	= - \dfrac{1}{3} \sum_{j=1,2,3} \int \diff{}{\vec{x}}
		\expval{V^{\alpha}_{j}(t,\vec{x})
                  V^{\Pphoton}_{j}(0)}_\text{\acs{qcd}}, \quad\alpha=\PZ,\,\Pphoton,
\end{gathered}
\end{equation}
where $V^{\Pphoton}$ is the electromagnetic current and $V^{\PZ}$ is the vector component of the $\PZ$ boson current.
Since at $\order{a}$ the breaking of global $\text{SU}(N_f)$ symmetry by the quark mass matrix leads to a mixing between the local currents of different quark flavours, we decide to work in the isospin basis where renormalisation and $\order{a}$ improvement are more easily implemented \cite{Gerardin:2018kpy},
\begin{equation}
\begin{gathered}
	V^{\Pphoton} = V^{3} + 1/\sqrt{3} V^{8} + 4/9 V^{\Pcharm},
	\\
	V^{\PZ} = (1/2-\weak(0)) V^{\Pphoton} - 1/6 V^{0} - 1/12 V^{\Pcharm}.
\end{gathered}
\end{equation}
Here, the isovector $V^{3}$, isoscalar $V^{8}$ and isosinglet $V^{0}$
components can be expressed in terms of the quark currents as
\begin{align}
	V^{3} &= \dfrac{1}{2} \left(V^{\Pup} - V^{\Pdown}\right),&
	V^{8} &= \dfrac{1}{2\sqrt{3}} \left(V^{\Pup} + V^{\Pdown} - 2V^{\Pstrange}\right),&
	V^{0} &= \dfrac{1}{2}\left(V^{\Pup} + V^{\Pdown} + V^{\Pstrange}\right),
\end{align}
while $V^c$ denotes the charm quark contribution. For the
single-flavour quark-bilinears at the sink, we use either the
\ac{local} or \ac{ps} discretisations, while we always use the
\ac{local} case at the source. The kernel $K(t,Q^2) = t^2 - 4/Q^2
\sin^2 (Qt/2)$ can, in principle, be evaluated at any $Q^2$. However,
in practice we are limited by the lattice spacing and the box size,
which are the cutoffs of our theory. In particular, choosing $Q^2 \sim
(\pi/\uvcut)^2$ probes the correlator at short distances, where one
finds strong discretisation effects. By contrast, $Q^2 \ll
\SI{1}{\giga\eV\squared}$ corresponds to the long-distance part of the
correlator, which is noisier and suffers from stronger finite-size
effects.

\section{Analysis}

\begin{figure}
		\centering
		\begin{subfigure}{0.42\textwidth}
			\centering
			\includegraphics[width=\textwidth]{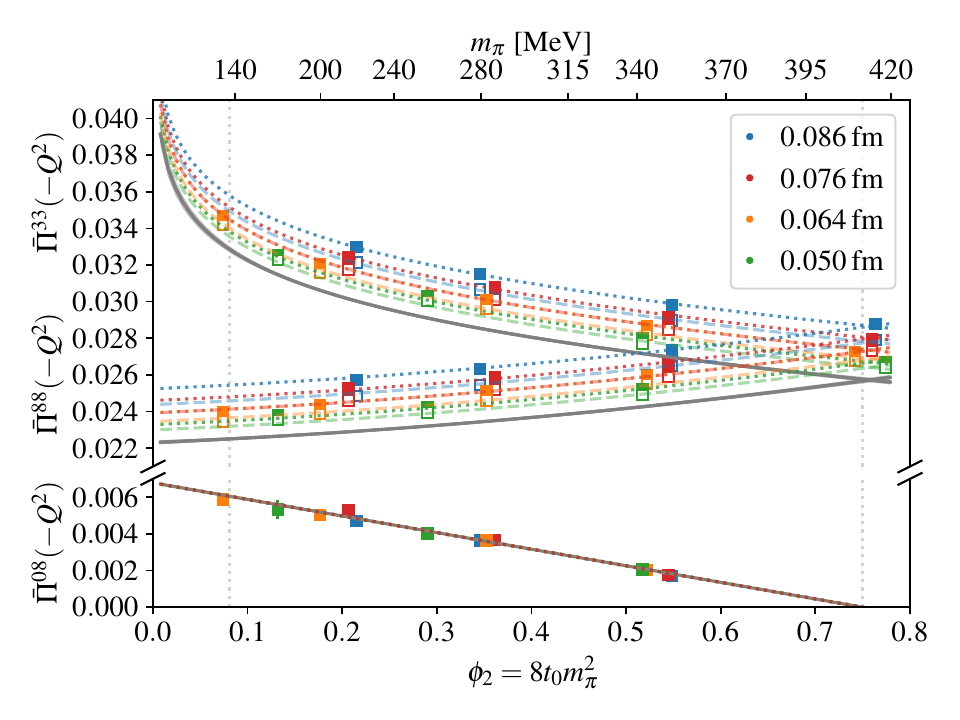}
		\end{subfigure}
		\begin{subfigure}{0.42\textwidth}
			\centering
			\includegraphics[width=\textwidth]{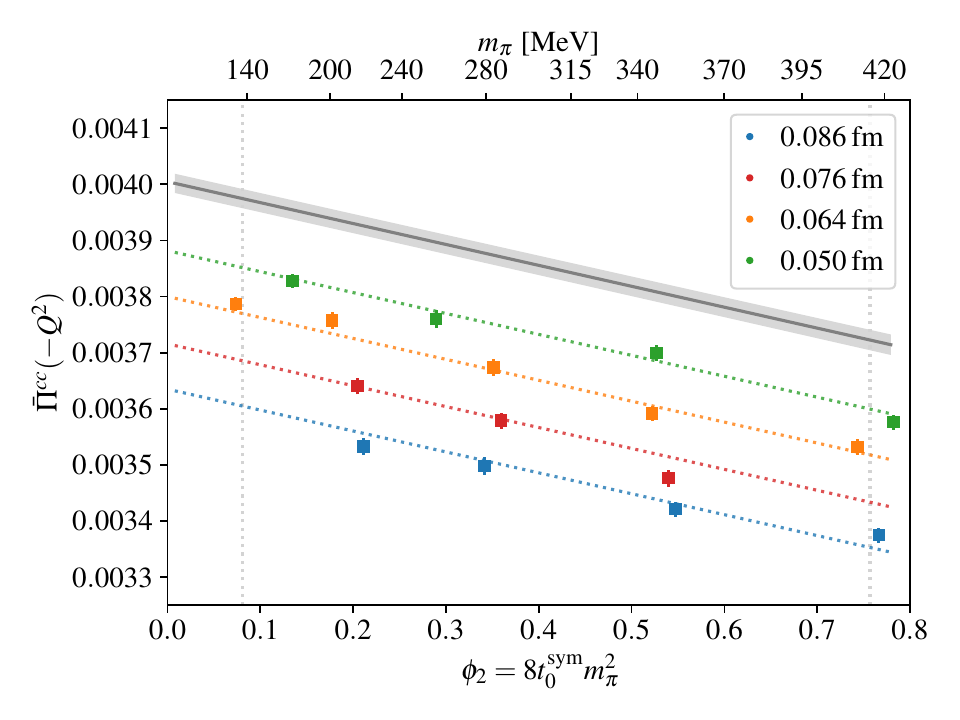}
		\end{subfigure}	
		\caption{Extrapolation to the physical point at $Q^2 =
                  \SI{1}{\giga\eV\squared}$. Left:
                  $\svpf^{33},\,\svpf^{88}$ and $\svpf^{08}$
                  contributions, incorporating the constraint
                  $\svpf^{33}=\svpf^{88}$ at the SU(3)-symmetric
                  point. Right: Extrapolation of the charm quark
                  contribution $\svpf^c$.}
		\label{fig:extrapolation-example}
	\end{figure}

\begin{figure}
	\centering
	\includegraphics[width=0.8\textwidth]{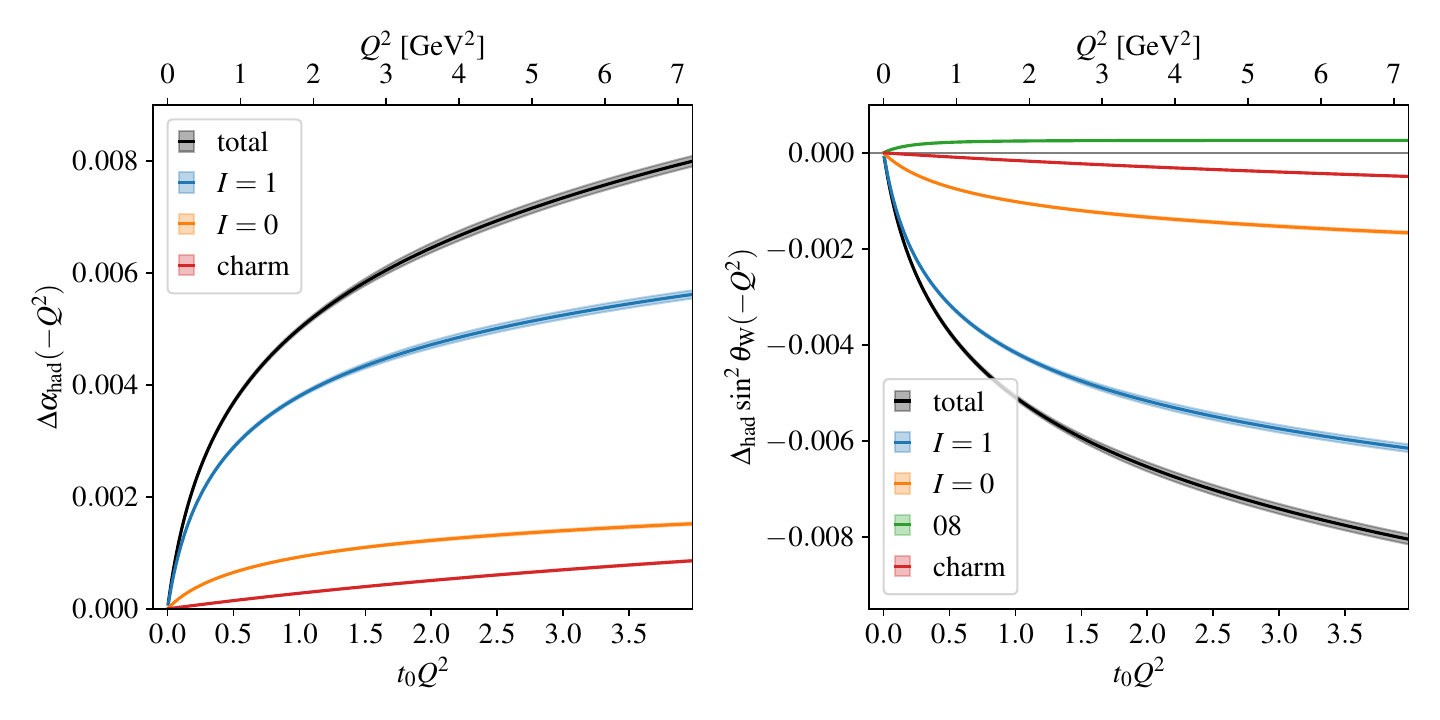}
	\caption{Leading hadronic contribution to the \acs{qed} coupling
		and electroweak mixing angle, which appear in black.
		We also give the various components: charm $G^{\Pcharm\Pcharm}$ in red,
		isoscalar $G^{88}$ in orange, isovector $G^{33}$ in blue, and
		mixing $G^{08}$ in green.
		The band width indicates the total error.
	}
	\label{fig:running}
\end{figure}

% Bounding method
To improve the signal-to-noise ratio for the vector correlators
$G^{\gamma\gamma}(t)$ and $G^{\PZ\gamma}(t)$ at large times~$t$, we
apply the bounding method \cite{Budapest-Marseille-Wuppertal:2017okr},
using the effective mass for the correlator's lower bound, and the
ground-state energy for the upper bound. For the isovector component,
the latter corresponds to the $\Prho$ meson or the two-pion state,
depending on the pion mass for a given ensemble. For the isoscalar,
the ground level is either the three-pion state or the $\omega$
meson. It is also possible to bound $\svpf^{08}$ using the effective
mass as upper bound and the isoscalar ground state as lower bound
\cite{Ce:2022eix,20.500.12030_7129}.
% Finite-size correction
To obtain the correlators in infinite volume, we decompose the
two-point functions in finite volume according to $G(t,\infty) =
G(t,\ircut) + \Delta G(t,\ircut)$, where $G(t,\ircut)$ is the
correlator computed on the lattice in finite volume, and $\Delta
G(t,\ircut)$ denotes the correction for \ac{fse} on a given timeslice,
which depend on the \acs{ir} regulator $\ircut$. To estimate $\Delta
G(t,\ircut)$, we have used the \ac{mll} method
\cite{Meyer:2011um,Lellouch:2000pv,Luscher:1991cf} and the \ac{hp}
procedure \cite{Hansen:2019rbh,Hansen:2020whp}. Both methods produce
consistent estimates for the finite-size correction, and we find that
$\Delta G(t,\ircut)$ amounts to a $\sim 2\%$ upward shift in the
isovector component $\svpf^{33}(-\SI{1}{\giga\eV\squared})$ at
$m_{\Ppi}\ircut = 4$, while the effect is reduced to $\sim 0.2\%$ at
$m_{\Ppi}\ircut = 6$. In addition, we have two sets of ensembles with
the same parameters but different volumes, as can be seen in
\Cref{fig:ensembles}, and we observe good agreement between them once
the \acs{fse} have been applied.
% Extrapolation to the physical point
Afterwards, we combine the extrapolation to the continuum $\uvcut
\rightarrow 0 $ and the interpolation to the isospin-symmetric pion
$m_{\Ppi}^\text{phy} = \SI{134.9768(5)}{\mega\eV}$ and kaon masses
$m_{\PK}^\text{phy} = \SI{495.011(15)}{\mega\eV}$
\cite{Risch:2021hty,Zyla:2020zbs}. To this end, we employ the
dimensionless fit variables $a^2/t_0^\text{sym}$, $\phi_2 = 8t_0
m_{\Ppi}^2$, and $\phi_4 = 8t_0 (m_{\Ppi}^2/2 + m_{\PK}^2)$, where
$t_0$ is measured on each ensemble, and $t_0^\text{sym}$ is taken at
the symmetric point $m_{\Ppi} = m_{\PK}$ from
\cite{Bruno:2016plf}. \Cref{fig:extrapolation-example} shows the fit
at $Q^2 = \SI{1}{\giga\eV\squared}$.
% 08 extrapolation
For $\svpf^{08}$, we only have \acs{ps}\acs{local}-data, but one
discretisation is sufficient as we do not discern any lattice spacing
dependence. By performing the quark contractions, we infer that
$\svpf^{08}=0$ whenever $m_{\Ppi} = m_{\PK}$, so our fit model must be
proportional to the combination $\phi_4 - 3/2\phi_2$. In fact, a
single parameter is enough to fit the data with $\chi^2/\text{dof}
\sim 0.7$,
\begin{equation}
	\label{eq:model-08}
	\svpf^{08,\ps\local}_{\text{model}} =
		\lambda_1 \left(\phi_4 - 3/2\phi_2\right).
\end{equation}
% cc extrapolation
For the $\svpf^{\Pcharm\Pcharm}$ contribution, we decide to only use
the \acs{ps}\acs{local}-data, which have $\order{10\%}$ discretisation
effects, and drop the \acs{local}\acs{local}-data, for which these
effects are as large as $\order{40\%}$. Replacing $t_0$ by
$t_0^\text{sym}$ in the \acs{tmr} kernel and the x-variable $\phi_2$,
the pion mass behaviour can be modelled using a linear term. This
substitution introduces correlations among all ensembles at the same
lattice spacing \cite{Bruno:2016plf}, which increases the size of the
covariance matrix in our fit. Therefore, we decide to fit the charm
contribution separately. Empirically, we find that an $a^2$-term is
sufficient to describe lattice artefacts, and hence we fit the Ansatz
\begin{equation}
	\label{eq:model-cc}
	\svpf^{\Pcharm\Pcharm}_{\text{con},\text{model}} =
		\svpf_{\text{con}}^{\Pcharm\Pcharm,\text{sym}}
		+ \delta_2^{\Pcharm\Pcharm,\ps\local}a^2/t_0^{\text{sym}}
		+ \gamma_{1}^{\Pcharm\Pcharm} (\phi_2 - \phi_2^{\text{sym}}).
\end{equation}
The subscript ``$\text{con}$'' indicates we only compute the quark-connected component.
Fitting the entire set of ensembles, we obtain $\chi^2/\text{dof} \sim 3$, but this is probably a side-effect from the increased size of the covariance matrix. Removing the ensembles with $m_{\Ppi} > \SI{400}{\mega\eV}$, we obtain $\chi^2/\text{dof} \sim 2$, and further removing ensembles with $m_{\Ppi} > \SI{300}{\mega\eV}$ yields $\chi^2/\text{dof} \sim 0.8$. The cuts have a negligible effect on the expectation value, and the quality of the fit barely depends on $Q^2$ \cite{20.500.12030_7129}.
% 33 extrapolation
Moving on to the fit function for the $\ps\local$ isovector component
$\svpf^{33}$, we make the ansatz
\begin{equation}
	\label{eq:isovector-model}
	\svpf^{33,\ps\local} =
		\svpf^{\text{sym}} + \delta_2^{\ps} a^2/t_0^{\text{sym}}
		+ \gamma_{1}^{33} \left(\phi_2 - \phi_2^{\text{sym}}\right)
		+ \gamma_{2}^{33} \log(\phi_2/\phi_2^{\text{sym}})
		+ \eta_1 (\phi_4 - \phi_4^{\text{sym}}).
\end{equation}
The $\gamma_{1}^{33}$ term models the dependence at large pion masses,
while \acs{chpt} inspires the form of the $\gamma_{2}^{33}$ term to
model the singular behaviour towards $m_{\Ppi} \to 0$
\cite{Golterman:2017njs}. Regarding the parameter $\eta_1$, we note
that our ensembles fulfil $\phi_4 \approx \text{const}$, and the
deviations from the exact equality can be modelled using a linear
term. We also note that the fit parameters $\svpf^{\text{sym}}$,
$\phi_2^{\text{sym}}$ and $\phi_4^{\text{sym}} =
3/2\phi_2^{\text{sym}}$ determine the coordinates where $m_{\Ppi} =
m_{\PK}$ and, since $\svpf^{33} = \svpf^{88}$ at this point, these set
of parameters are common for both isospin channels.
% 88 extrapolation
To model the $\ps\local$ isoscalar component, we use
\begin{equation}
	\label{eq:isoscalar-model}
	\svpf^{88,\ps\local} =
		\svpf^{\text{sym}} + \delta_2^{\ps\local} a^2/t_0^{\text{sym}}
		+ \gamma_{1}^{88} \left(\phi_2 - \phi_2^{\text{sym}}\right)
		+ \gamma_{2}^{88} (\phi_2 - \phi_2^{\text{sym}})^2
		+ \eta_1 (\phi_4 - \phi_4^{\text{sym}}).
\end{equation}
Similar expressions are used for the \acs{local}\acs{local}-data. In
this case, the model has a finite limit towards $m_{\Ppi} \to 0$, as
expected in \acs{chpt} \cite{Golterman:2017njs}, although the
particular form for both isospin channels is chosen such that the fit
faithfully describes the result obtained on the ensemble at the
physical pion mass (E250). Using
\cref{eq:isovector-model,eq:isoscalar-model}, we obtain
$\chi^2/\text{dof} \sim 1.5$ up to $Q^2 \gtrsim
\SI{2.5}{\giga\eV\squared}$. Beyond this point, the increasing size of
lattice artefacts requires the inclusion of an extra
$(a^2/t_0^{\text{sym}})^{3/2}$ term in both
\cref{eq:isovector-model,eq:isoscalar-model}, allowing us to reach
$Q^2 \sim \SI{7}{\giga\eV\squared}$ with similar fit quality, albeit
increasing the statistical error. We effect the transition between
both models using a smooth step function centred around
$\SI{2.5}{\giga\eV\squared}$.
% Error budget
At the physical point, with $Q^2 = \SI{1}{\giga\eV\squared}$, we obtain
\begin{equation}
\begin{gathered}
	\qedchad(\SI{-1}{\giga\eV\squared})\times 10^6 = 3864~(17)~(8)~(22)~(4)~(12)~[32,~\SI{0.8}{\%}],\\
	\weakhad(\SI{-1}{\giga\eV\squared})\times 10^6 = -3927~(19)~(5)~(32)~(4)~(13)~[40,~\SI{1.0}{\%}],
\end{gathered}
\end{equation}
where the errors from left to right are from statistics,
extrapolation, scale-setting, missing charm-quark loops and
\acs{ib}. In square brackets, we add all errors in quadrature. The
last number on the right shows that we obtain a precision of $\sim
1\%$ at this particular momentum. The statistical error is propagated
using bootstrap sampling, and the extrapolation uncertainty is
obtained by repeating the fit removing the heavier pion masses. Note
that the calibration of the lattice scale $\sqrt{8t_0} =
0.415~(4)~(2)~\text{fm}$ \cite{Bruno:2016plf} enters
indirectly in our analysis ($\svpf$ is dimensionless) through the
\acs{tmr} kernel and the definition of $\phi_2$ and $\phi_4$, and we
estimate the final uncertainty propagating the error of $t_0$ using
bootstrap sampling. We find ourselves in a favourable position,
because $\sqrt{8t_0}$ is determined with $1\%$ accuracy and it induces
a $\sim 0.7\%$ error in the final quantities. Nonetheless, the
scale-setting uncertainty is dominant in the range $0 < Q^2 \lesssim
\SI{3}{\giga\eV\squared}$. As a result, there is an ongoing effort to
improve the determination of the scale, including \acs{ib} effects
\cite{Segner:2021yqo}. Regarding the missing charm-quark contribution
to the quark sea, we estimate the charm quenching effect
phenomenologically, quantifying the contributions from
$\PDplus\PDminus$, $\PDzero\APDzero$ and $\PDsplus\PDsminus$ to the
\acs{hvp} treating the $\PD$-meson form factors in scalar \acs{qed}.
Besides, the valence charm-quark loops are negligible according to
\cite{Budapest-Marseille-Wuppertal:2017okr}. Finally, we have
evaluated the quark-connected \acs{hvp} in $\acs{qcd}+\acs{qed}$ on
one ensemble at $m_{\Ppi} \sim \SI{220}{\mega\eV}$. The result is
used to estimate the relative size of the missing \acs{ib} effects
that we add to our error budget at the physical point.
%There is an ongoing program in the Mainz group to extend this calculation to more ensembles and perform the extrapolation to the physical point \cite{20.500.12030_6324}.
% Q^2 dependence
We repeat the extrapolation at several $Q^2$, distributed
logarithmically, to probe the low-momentum region, and we plot the
results in \Cref{fig:running}.
% Pade approximations
To provide the running in an analytic form, we use the fact that the
\acs{hvp} momentum dependence can be written as a Stieltjes function
\cite{Aubin:2012me}. In turn, this can be approximated by a ratio of
polynomials, i.e. a Pad\'{e} approximant $R^N_M(Q^2)$, whose general
expression is
\begin{equation}
	R^N_M(Q^2) = \dfrac{\sum_{j=0}^{M} a_{j} Q^{2j}}{1 + \sum_{k=1}^{N} b_{k} Q^{2k}}.
\end{equation}
Via a least-squares fit of $R^N_M(Q^2)$ to $\svpf(-Q^2)$, we obtain
\begin{equation}
	\label{eq:pade-approximants}
	\begin{gathered}
	\qedchad(-Q^2) =
	4\pi\qedc~\dfrac{\SI{0.1094(23)}{x} + \SI{0.093(15)}{x^2} + \SI{0.0039(6)}{x^3}}{1 + \SI{2.85(22)}{x} + \SI{1.03(19)}{x^2} + \SI{0.0166(12)}{x^3}},
	\\
	\weakhad(-Q^2) =
	-\dfrac{4\pi\alpha}{\weak}~\dfrac{\SI{0.02263(6)}{x} + \SI{0.025(5)}{x^2} + \SI{0.00089(34)}{x^3}}{1 + \SI{2.94(29)}{x} + \SI{1.12(27)}{x^2} + \SI{0.015(8)}{x^3}},
	\end{gathered}
\end{equation}
with $\text{x}=Q^2/\text{GeV}^2$, $4\pi\qedc =
\num{0.091701236853(14)}$ and $\weak = \num{0.23857(5)}$
\cite{Zyla:2020zbs}. Note the parameter $a_0$ is zero since
$\svpf(0) = 0$. The choice $N=M=3$ reproduces the data accurately,
yet we observe that the extra fit parameters are poorly determined. In
order to reproduce the error bands in \Cref{fig:running}, we refer the
reader to \cite{Ce:2022eix}, where we also include the correlation
matrix of the fit parameters.

\begin{figure}
  \centering
  \includegraphics[width=0.6\textwidth]{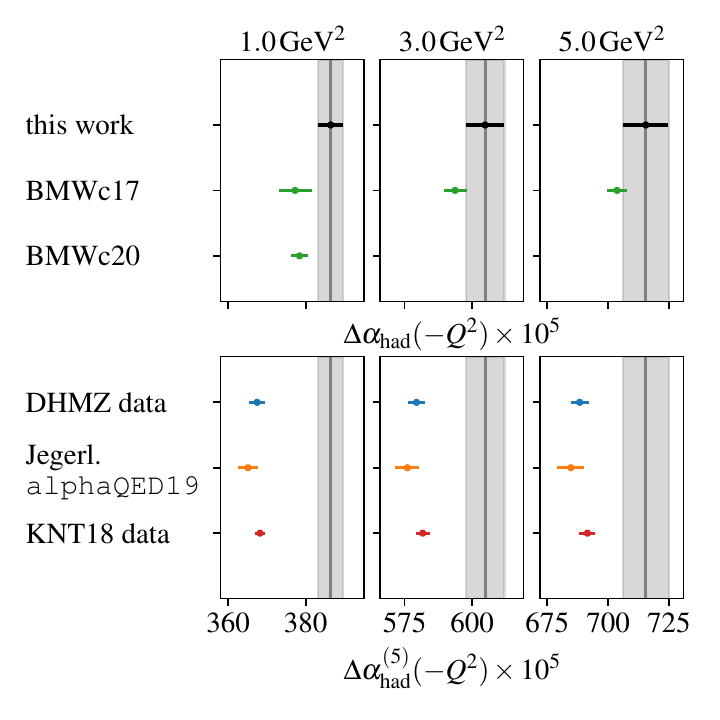}
  \caption{Comparison of the \acs{qed} coupling at various momenta
    between our determination and several others. At the top half, the
    lattice determinations by \acs{bmwc}
    \cite{Borsanyi:2020mff,Budapest-Marseille-Wuppertal:2017okr} and,
    at the bottom, the 5 flavours comparison with the phenomenological
    determinations by \acs{dhmz} \cite{Davier:2019can}, Jegerlehner
    \cite{Jegerlehner:2019lxt}, and \acs{knt}
    \cite{Keshavarzi:2019abf}.}
\label{fig:comparison}
\end{figure}

\section{The hadronic running of $\alpha$ at the $Z$-pole}

The value of $\Delta\alpha_{\rm had}^{(5)}(q^2)$ at $q^2=M_Z^2$ is a
key quantity in electroweak precision physics. It serves, on the one
hand, as an input quantity for the interpretation of experiments at
high-energy colliders. On the other hand, direct theoretical
determinations can be compared to the results of global electroweak
fits, thereby providing a stringent test of the SM.

The traditional method to determine $\Delta\alpha_{\rm
  had}^{(5)}(M_Z^2)$ proceeds by evaluating a dispersion integral over
the hadronic cross section ratio (``$R$-ratio'') $R(s)$, according to
\begin{equation}\label{eq:Delta-alpha-disp}
  \Delta\alpha_{\rm had}^{(5)}(q^2) = -\frac{\alpha\,q^2}{3\pi}
  {\cal P}\hspace{-12pt}\int_{m_{\pi_0}^2}^{\infty}\,\frac{R(s)}{s(s-q^2)}\,ds,\quad
  q^2=M_Z^2,\quad
  R(s)=\frac{3s}{4\pi\alpha(s)}\,\sigma(e^+e^-\to\hbox{hadrons})\,. 
\end{equation}
The above master formula is closely related to the corresponding
dispersion integral for the leading-order hadronic vacuum polarisation
contribution to the muon $g-2$, i.e.
\begin{equation}\label{eq:amu-disp}
  a_\mu^{\rm LO,\,hvp}=\left(\frac{\alpha m_\mu}{3\pi}\right)^2
  \int_{m_{\pi_0}^2}^{\infty}\,\frac{R(s)\,\hat{K}(s)}{s^2}\,ds,\quad
  0.63\leq\hat{K}(s)\leq1\,. 
\end{equation}
This implies that the evaluation of $\Delta\alpha_{\rm
  had}^{(5)}(M_Z^2)$ by means of eq.\,(\ref{eq:Delta-alpha-disp}) is
affected by experimental uncertainties arising from experimentally
measured hadronic cross sections in a similar manner than data-driven
determinations of $a_\mu^{\rm LO,\,hvp}$.
%{{\bf Comment:} include statement on different kernel functions?}

An alternative approach to evaluate $\Delta\alpha_{\rm
  had}^{(5)}(M_Z^2)$ is based on the so-called Euclidean split
technique (also dubbed the Adler function approach)
\cite{Eidelman:1998vc, Jegerlehner:1999hg, Jegerlehner:2008rs}, in
which $\Delta\alpha_{\rm had}^{(5)}(M_Z^2)$ is divided into three
separate contributions, according to
\begin{equation}\label{eq:EuclSplit}
  \Delta\alpha_{\rm had}^{(5)}(M_Z^2) =
  \Delta\alpha_{\rm had}^{(5)}(-Q_0^2) +
   \Big[\Delta\alpha_{\rm had}^{(5)}(-M_Z^2)
   -\Delta\alpha_{\rm had}^{(5)}(-Q_0^2)\Big] +
   \Big[\Delta\alpha_{\rm had}^{(5)}(M_Z^2)
   -\Delta\alpha_{\rm had}^{(5)}(-M_Z^2)\Big].
\end{equation}
It is primarily the first term on the right-hand side,
$\Delta\alpha_{\rm had}^{(5)}(-Q_0^2)$, that absorbs the bulk of the
non-perturbative physics, depending on the choice of the Euclidean
squared momentum transfer $Q_0^2$. In particular, the first term in
square brackets on the RHS of eq.\,(\ref{eq:EuclSplit}) can be
computed as an integral over the Adler function $D(Q^2)$, defined by
\begin{equation}\label{eq:Adlerdef}
  D(-s):=\frac{3\pi}{\alpha}\,s\,\frac{d}{ds}\Delta\alpha_{\rm had}(s)\,,
\end{equation}
and which is known in massive QCD perturbation theory at three loops
\cite{Chetyrkin:1996cf, Eidelman:1998vc,
  Jegerlehner:2008rs}. Integrating eq.\,(\ref{eq:Adlerdef}) for
spacelike momentum transfers from $Q_0^2$ to $M_Z^2$ yields
\begin{equation}\label{eq:running_Adler}
  \Big[\Delta\alpha_{\rm had}^{(5)}(-M_Z^2)
    -\Delta\alpha_{\rm had}^{(5)}(-Q_0^2)\Big]_{\rm pQCD/Adler} =
  \frac{\alpha}{3\pi}\int_{Q_0^2}^{M_Z^2}\,\frac{dQ^2}{Q^2}\,D(Q^2).
\end{equation}
Furthermore, by inserting the dispersion integral for
$\Delta\alpha_{\rm had}$ into eq.\,(\ref{eq:Adlerdef}) one obtains a
representation of $D(Q^2)$ in terms of the $R$-ratio, i.e.
\begin{equation}
  D(Q^2)=Q^2\int_{m_{\pi_0^2}}^{\infty}\,\frac{R(s)}{(s+Q^2)^2}\,ds\,,
\end{equation}
and a straightforward calculation shows that the second term on the
RHS of eq.\,(\ref{eq:EuclSplit}) can also be expressed as
\begin{equation}\label{eq:running_disp}
  \Big[\Delta\alpha_{\rm had}^{(5)}(-M_Z^2)
   -\Delta\alpha_{\rm had}^{(5)}(-Q_0^2)\Big]_{\rm disp.} =
  \frac{\alpha(M_Z^2-Q_0^2)}{3\pi}
  \int_{m_{\pi_0^2}}^{\infty}\,\frac{R(s)}{(s+Q_0^2)(s+M_Z^2)}ds\,.
\end{equation}
The freedom to evaluate this quantity either in perturbative QCD or in
terms of the experimentally mesured $R$-ratio allows for a valuable
cross check. Finally, the third term on the RHS of
eq.\,(\ref{eq:EuclSplit}) provides the link between spacelike and
timelike regimes, which, at energies as large as the $Z$~boson mass,
can be reliably determined in perturbation theory
\cite{Jegerlehner:1985gq, Jegerlehner:2019lxt}, viz.
\begin{equation}
  \Big[\Delta\alpha_{\rm had}^{(5)}(M_Z^2)
   -\Delta\alpha_{\rm had}^{(5)}(-M_Z^2)\Big] = 0.000\,045(2)\,.
\end{equation}
The Euclidean split technique holds several advantages over the
standard method based on dispersion integrals:
\begin{itemize}
  \item The integration in eq.\,(\ref{eq:running_Adler}) is performed
    over Euclidean squared momenta, and therefore the integration over
    resonances and physical thresholds, which renders the evaluation
    of the dispersion integrals in eqs.\,(\ref{eq:Delta-alpha-disp})
    and\,(\ref{eq:amu-disp}) quite intricate, can be avoided.
  \item The non-perturbative threshold value $\Delta\alpha_{\rm
    had}^{(5)}(-Q_0^2)$ can be determined either via dispersion theory
    and the experimental $R$-ratio or in lattice QCD. In the future,
    there is also the possibility of a direct experimental measurement
    from the MUonE experiment \cite{Venanzoni:2018ktr,
      Masiero:2020vxk}. By contrast, the direct evaluation of
    $\Delta\alpha_{\rm had}^{(5)}(M_Z^2)$ via dispersion integrals
    requires precise experimental data up to much higher energies
    compared to $\Delta\alpha_{\rm had}^{(5)}(-Q_0^2)$.
\end{itemize}

\section{Estimate of $\Delta\alpha_{\rm had}^{(5)}(M_Z^2)$ from lattice QCD}

\begin{figure}
  \centering
  \includegraphics[width=0.8\textwidth]{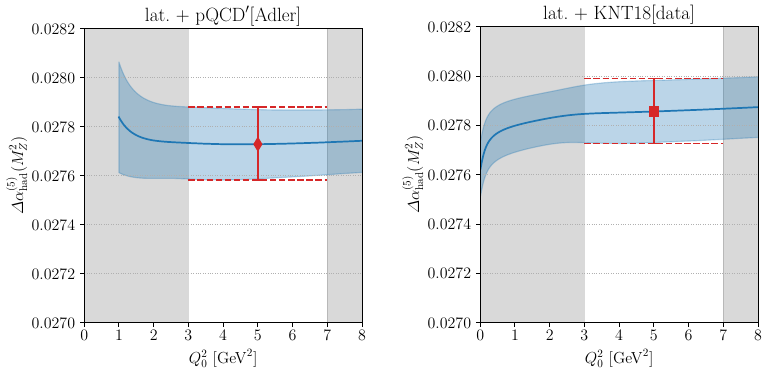}
  \caption{Left: The hadronic running of $\alpha$ at the $Z$~pole,
    evaluated via the Euclidean split techniques and
    eq.\,(\ref{eq:running_Adler}), plotted as a function of the
    threshold scale, $Q_0^2$. Right: the same quantity with the
    running evaluated via dispersion theory accoring to
    eq.\,(\ref{eq:running_disp}), using the $R$-ratio data and
    covariance matrix from
    \cite{Keshavarzi:2018mgv,Keshavarzi:KNT18data}. The red point in
    both panels represents the final estimate, while the horizontal
    dashed lines denote the uncertainty due to the choice of $Q_0^2$
    as inferred from the maximum and minimum values within the
    interval from~3 to~7 GeV$^2$.}
  \label{fig:Delta-alpha}
\end{figure}

In order to produce an estimate for $\Delta\alpha_{\rm
  had}^{(5)}(M_Z^2)$, we substitute our lattice results obtained at
small Euclidean momentum transfers for the offset value
$\Delta\alpha_{\rm had}^{(5)}(-Q_0^2)$ in
eq.\,(\ref{eq:EuclSplit}). Furthermore, we employ the Adler function
to determine the running from low to high Euclidean momenta, by
computing $[\Delta\alpha_{\rm had}^{(5)}(-M_Z^2) -\Delta\alpha_{\rm
    had}^{(5)}(-Q_0^2)]_{\rm pQCD/Adler}$ using the software package
{\tt pQCD/Adler} by Jegerlehner \cite{Jegerlehner:pQCDAdler}. In order
to assess the uncertainty due to the ambiguity in the choice of the
threshold scale $Q_0^2$, we plot in Fig.\,\ref{fig:Delta-alpha} the
resulting estimates for $\Delta\alpha_{\rm had}^{(5)}(M_Z^2)$ as a
function of $Q_0^2$. We observe stability in the estimates for
$\Delta\alpha_{\rm had}^{(5)}(M_Z^2)$ when $Q_0^2$ is varied between~3
and~7\,GeV$^2$. As a cross check, we have determined the running using
dispersion theory (see eq.\,(\ref{eq:running_disp})), and the
resulting estimates for $\Delta\alpha_{\rm had}^{(5)}(M_Z^2)$ are
again plotted versus $Q_0^2$ in the right panel of
Fig.\,\ref{fig:Delta-alpha}. We find that both alternatives yield very
compatible results within errors. As our final estimate we quote the
result based on the integration of the perturbative Adler function,
i.e.
\begin{equation}\label{eq:final}
  \Delta\alpha_{\rm had}^{(5)}(M_Z^2)=0.027\,73(9)_{\rm lat}(2)_{\rm
    btm}(12)_{\rm pQCD}\,[15]_{\rm tot}\,,
\end{equation}
where the first error is the intrinsic error of our lattice
calculation, including the ambiguity in the choice of $Q_0^2$, the
second is an estimate of the uncertainty due to the missing bottom
quark contribution, and the last error arises from the running from
$Q_0^2$ to $M_Z^2$ is evaluated in terms of the integrated Adler
function computed in perturbative QCD. The final number in square
brackets denotes the total error after summing the individual
uncertainties in quadrature.

\begin{figure}[t]
  \centering
  \includegraphics[width=0.99\textwidth]{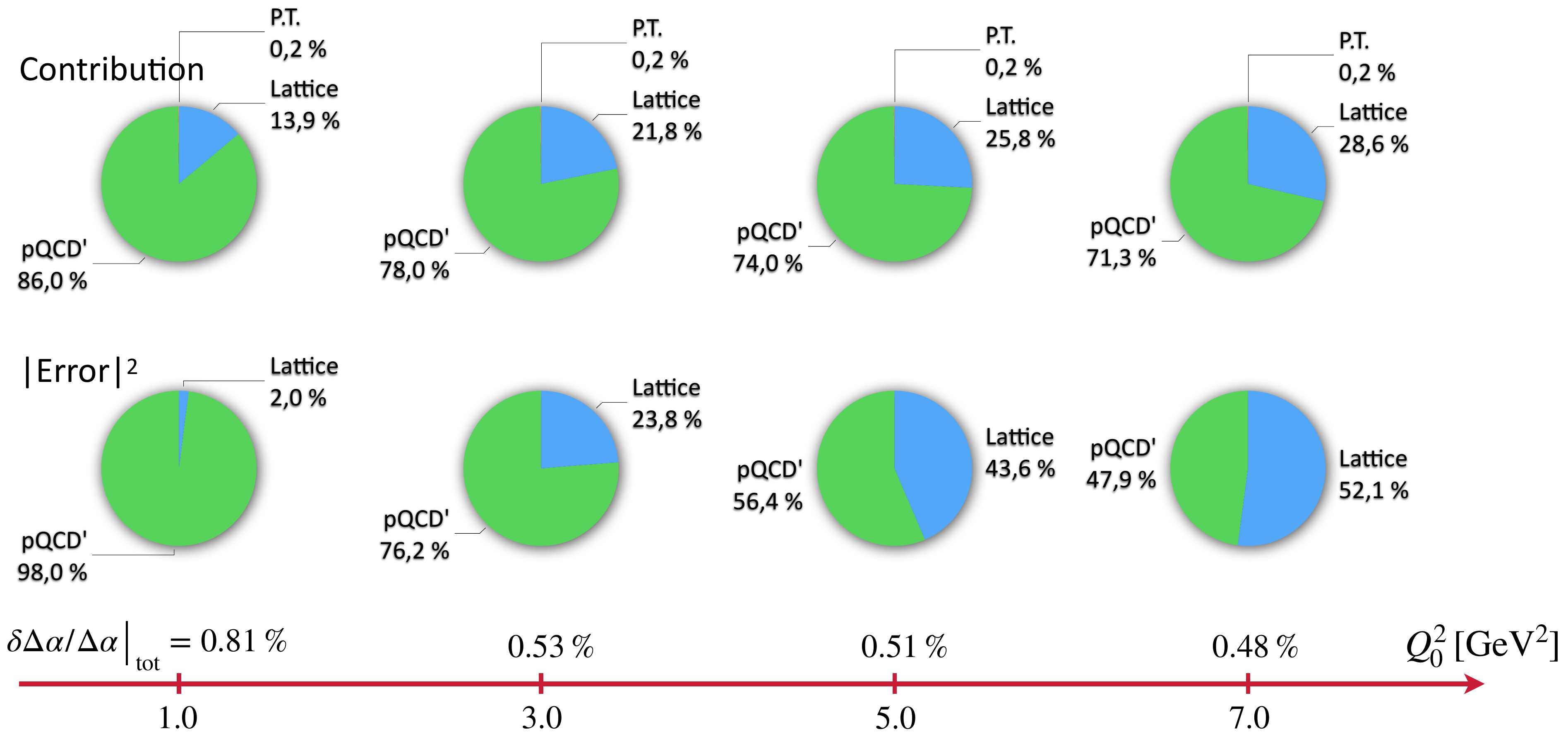}
  \caption{Pie charts showing the relative contribution to
    $\Delta\alpha_{\rm had}^{(5)}(M_Z^2)$ from the Euclidean split
    technique, for several values of $Q_0^2$. The running between
    $Q_0^2$ and $M_Z^2$ is evaluated using the perturbative Adler
    function. The top and bottom rows show the contributions to the
    value and the variance, respectively.}
  \label{fig:piecharts}
\end{figure}

It is instructive to study the relative size of the individual
contributions in the Euclidean split technique,
eq.\,(\ref{eq:EuclSplit}), as a function of $Q_0^2$. This is
illustrated by the pie charts in Fig.\,\ref{fig:piecharts}, where the
top row shows the relative size of the individual terms that make up
the central value of $\Delta\alpha_{\rm had}^{(5)}(M_Z^2)$, while the
charts in the bottom row represent their contributions to the
variance. This exercise shows that the scale $Q_0^2$ can be used to
optimise the reliability and precision of $\Delta\alpha_{\rm
  had}^{(5)}(M_Z^2)$. For instance, a significant reduction of the
total error in the lattice calculation will do little to improve the
overall precision of $\Delta\alpha_{\rm had}^{(5)}(M_Z^2)$ if the
threshold scale is fixed at 3\,GeV$^2$. From the charts in the figure
one reads off that our lattice calculation accounts for $\sim25\%$ of
the value of the hadronic running and for $(25-50)\%$ of the variance,
depending on the value of $Q_0^2$ in the interval between~3 and
7~GeV$^2$.

\begin{figure}[t]
  \centering
  \includegraphics[width=0.6\textwidth]{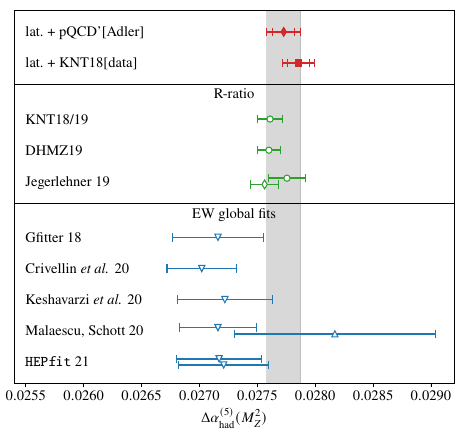}
  \caption{Compilation of results for $\Delta\alpha_{\rm
      had}^{(5)}(M_Z^2)$. Our lattice estimates are shown as red
    symbols, with our preferred result, listed in
    eq.\,(\ref{eq:final}) shown as the grey vertical band. Green
    circles represent results based on dispersion theory
    \cite{Keshavarzi:2018mgv, Keshavarzi:2019abf, Davier:2019can,
      Jegerlehner:2019lxt}, where the $R$-ratio integration is
    performed over the entire momentum range. Jegerlehner's estimate
    applying the Euclidean split technique with dispersive input for
    $\Delta\alpha_{\rm had}(-Q_0^2)$ is shown as the green open
    diamond. Open blue downward pointing triangles show results from
    global electroweak fits \cite{Haller:2018nnx, Crivellin:2020zul,
      Keshavarzi:2020bfy, Malaescu:2020zuc, deBlas:2021wap} with
    $\Delta\alpha_{\rm had}(M_Z^2)$ as a free fit parameter. The
    upward pointing open triangle denotes a fit in which the Higgs
    mass is left as another free parameter.}
  \label{fig:compilation}
\end{figure}

In Fig.\,\ref{fig:compilation} we compare our results with other
direct estimates obtained via the data-driven approach and the results
from global electroweak fits. Our preferred result of
eq.\,(\ref{eq:final}) is shown as the grey vertical band. Within the
quoted errors, it agrees very well with the determinations based on
the $R$-ratio shown as green points in the middle panel. For instance,
Jegerlehner \cite{Jegerlehner:2019lxt} quotes a value of
$\Delta\alpha_{\rm had}^{(5)}(M_Z^2)=0.027\,52(12)$ using the
$R$-ratio for fixing the non-perturbative input at $Q_0^2=4\,\rm
GeV^2$ and the perturbative Adler function for the running. At first
sight, the agreement between the vertical grey band and the green
points in the middle panel of Fig.\,\ref{fig:compilation} appears to
contradict our earlier observation of a tension between lattice and
data-driven evaluations of $\Delta\alpha_{\rm had}(-Q_0^2)$ for
$Q_0^2\lesssim 7\rm GeV^2$ (see Fig.\,\ref{fig:comparison}). The
resolution of what seems like a contradiction comes from the
observation that the running from low to high Euclidean momenta is
correlated between the two approaches. In other words, both lattice
and data-driven determinations of $\Delta\alpha_{\rm
  had}^{(5)}(M_Z^2)$ share the correlated uncertainty in the
evaluation of $[\Delta\alpha_{\rm had}^{(5)}(-M_Z^2)
  -\Delta\alpha_{\rm had}^{(5)}(-Q_0^2)]$, which must be dropped when
computing the difference between the two methods.

In the bottom panel of Fig.\,\ref{fig:compilation} we plot various
results from global electroweak fits. Although the latter mostly
favour slightly smaller values for the hadronic running, the results
are not in contradiction with our lattice estimate, given the
relatively large errors.

\section{Conclusions}

We have presented results for the leading hadronic contribution to the
running of the electromagnetic coupling $\qedchad(-Q^2)$ and the
electroweak mixing angle $\weakhad(-Q^2)$ in the range of space-like
momenta $Q^2 \leq \SI{7}{\giga\eV\squared}$. We have estimated all
sources of uncertainty and find that the scale-setting error dominates
for $Q^2 \leq \SI{3}{\giga\eV\squared}$. For larger momenta, it is
necessary to include an extra $a^3$ term in the continuum
extrapolation, which increases the statistical error. Overall, we
achieve $1\%$ precision for both quantities in the region $Q^2 >
\SI{1}{\giga\eV\squared}$, and $2\%$ for smaller momenta. Our main
results are provided in the form of analytic functions for
$\qedchad(-Q^2)$ and $\weakhad(-Q^2)$, given in
\cref{eq:pade-approximants} in terms of Pad\'{e} Ans\"{a}tze.
Together with the corresponding correlation matrices for the
parameters $a_j$, $b_k$ \cite{Ce:2022eix}, it is possible to reproduce
our results and total uncertainty at any small space-like momentum
$Q^2$. Our result for $\Delta\alpha$ compares well with the lattice
determinations by the BMW collaboration
\cite{Borsanyi:2020mff,Budapest-Marseille-Wuppertal:2017okr}, with
only a mild tension of $1~\sigma$ to $2~\sigma$. However, we observe
a significant discrepancy of more than $3\sigma$ with the
phenomenological determinations by \acs{dhmz} \cite{Davier:2019can},
Jegerlehner \cite{Jegerlehner:2019lxt}, and \acs{knt}
\cite{Keshavarzi:2019abf}.

Given the close relation between the hadronic running of $\alpha$ and
the hadronic vacuum polarisation contribution to the muon $g-2$, the
observation of a tension in $\Delta\alpha$ between lattice and
data-driven estimates is consistent with the apparent discrepancy for
the intermediate window observable derived from $a_\mu^{\rm LO,\,hvp}$
\cite{Borsanyi:2020mff, Ce:2022kxy, Alexandrou:2022amy}. In spite of
the observed tension with data-driven approaches, we find that the
conversion of our lattice result for $\Delta\alpha_{\rm had}(-Q_0^2)$
into an estimate for $\Delta\alpha_{\rm had}^{(5)}(M_Z^2)$ broadly
agrees with global electroweak fits. Our calculation, therefore, is
not in contradiction with the SM, and we conclude that the SM can
accommodate a larger value for $a_\mu$ without producing a significant
tension with electroweak data, at least at the current level of
precision.

\section*{Acknowledgements}
Calculations for this project have been performed on the HPC clusters
Clover and HIMster-II at Helmholtz Institute Mainz and on Mogon-II at
Johannes Gutenberg-Universität (JGU) Mainz, on the HPC systems JUQUEEN
and JUWELS at J\"ulich Supercomputing Centre (JSC), and on Hazel Hen
at H\"ochstleistungsrechenzentrum Stuttgart (HLRS). The authors
gratefully acknowledge the support of the Gauss Centre for
Supercomputing (GCS) and the John von Neumann-Institut für Computing
(NIC) for project HMZ21 and HMZ23 at JSC and project GCS-HQCD at HLRS.
We are grateful to our colleagues in the CLS initiative for sharing
ensembles. This work has been supported by Deutsche
Forschungsgemeinschaft (German Research Foundation, DFG) through
project HI 2048/1-2 (project No.\ 399400745) and through the Cluster
of Excellence ``Precision Physics, Fundamental Interactions and
Structure of Matter'' (PRISMA+ EXC 2118/1), funded within the German
Excellence strategy (Project ID 39083149).
%The work of M.C.\ has been supported by the European Union's Horizon
%2020 research and innovation program under the Marie Skłodowska-Curie Grant Agreement No.\ 843134.
A.G.\ received funding from the Excellence Initiative of Aix-Marseille
University - A*MIDEX, a French ``Investissements d'Avenir'' programme,
AMX-18-ACE-005 and from the French National Research Agency under the
contract ANR-20-CE31-0016.

\bibliographystyle{JHEP}
\bibliography{bib}

\end{document}